# Confining Eutectic Gallium Indium (eGaIn) in Expired Artificial Kidneys to Unveil Nanoporous Conductive Wires

Momena Monwar[a], Gerra Licup[a], M. Rashed Khan[a,*]

[a]Department of Chemical and Materials Engineering, University of Nevada, Reno, NV-89557

*Corresponding Author: mrkhan@unr.edu

**Abstract**

Nanoporous membranes have gained considerable interest in drug delivery[1], ion transportation[2], micro/nanofluidics[3], molecular sensing[4], and separation science[5]. Artificial kidneys, also known as dialyzers, reject pathogens and other unwanted substances from the blood, utilize hundreds of soft and nanoporous polymeric microtubes, and slowly become a burden to the environment with the growing number of dialysis patients worldwide. We demonstrate the fabrication of nanoporous conductive wires utilizing empty polysulfone microtubes collected from expired and unused artificial kidneys, also known as medical wastes. Injecting a fluidic, highly conductive, and room temperature liquid alloy (eutectic gallium indium- eGaIn[6], 75% Ga, 25% In) into microtubes of a twenty years old dialyzer, here, we have revealed a new class of nanoporous and conductive functional materials. These conductive fibers upcycle a medical waste, do not require expensive and conventional fabrication processes, and still provide the quintessential metal-oxide/metal framework due to the presence of the native surface oxide (i.e., Gallium Oxide, $Ga_2O_3$) of eGaIn at the nanoconfinement (i.e., nanopores) for nano/biosensing. We harnessed these new materials to sense and differentiate microliter volumes of deionized (DI) water, 1M hydrochloric acid (HCl), and 95% ethanol (EtOH), leveraging their electrical signatures. This new class of soft nanomaterials has the potential to become the paradigm-shift platforms for the next-generation of biomedical, bioelectronics, nanoelectronics, and sensor devices.

**Main**

The demand for porous membranes in kidney dialyzers is continuously growing due to the increasing number of chronic kidney patients worldwide. Nearly 66% of the patients require high flux, highly permeable dialyzers to remove substances larger than Urea[7] immediately, and for the 3.2 million patients, when each patient requires an average of ~150 dialysis treatments, we require billions of high flux dialyzers worldwide[8]. In contrast, expired and unused dialyzers- commonly referred to as medical waste, raise environmental concerns before being incinerated impose additional concerns related to non-renewable resources, emission hazards, cost, and energy consumptions[8,9]. Additionally, hundreds of synthetic nanoporous microtubes within each dialyzer exaggerate these concerns even more. Therefore, hemodialyzers that remain unused and become expired need further investigation, and we assume that hundreds of porous and monolumen microtubes within the expired and unused dialyzers can be upcycled for functional applications. Focused electron/ion beams, controlled dielectric breakdown, laser pulling of glass or quartz pipette, electrochemical deposition, nanoimprinting, thermal annealing, chemical etching, hard masks, and metal catalysts are some of the processes to achieve nanoporous structures on silicon, quartz, metals, films, or 2D materials[10]. In contrast, nanoporous microtubes from expired dialyzers do not require any additional physical or chemical treatments/processes producing nanopatterns to control and manipulate fluids. We harnessed the nanoporous surface of the microtubes as the miniaturized nanoplatforms to diffuse fluids of interest to acquire distinct electrical signals through the interfacial interactions between the fabricated conductive wires and microdroplets of liquids.

This paper describes the fabrication of nanoporous conductive wires utilizing empty polysulfone microtubes collected from expired and unused artificial kidneys for the proof-of-concept, first-of-its-kind nano/biosensing applications. We retrieved empty hollow fibers from decades-old high-flux hemodialyzers commonly known as "artificial kidneys" and typically considered "medical waste" after the expiration. **Fig 1** demonstrates the topography, pore distribution, and mechanical strength of an empty nanoporous fiber collected from a twenty-year-old high-flux hemodialyzer. Under the optical microscope, the ~260μm diameter fiber appears smooth [**Fig 1a**], but the pores become visible under scanning electron micrographs (SEM) [**Fig 1b-d**]. **Fig 1e** shows the bar-chart of the pore distribution analyzed from **Fig 1d** using ImageJ software. The average pore size is ~ 770 nm, the highest pore diameter is ~2.04 μm, and the lowest is ~250 nm for this strand and other strands collected from the expired dialyzer. Each of these strands in the dialyzer effectively removes the solutes and water through the semipermeable membranes (polysulfone) using different mass separation mechanisms like diffusion, convection, and adsorption[11].

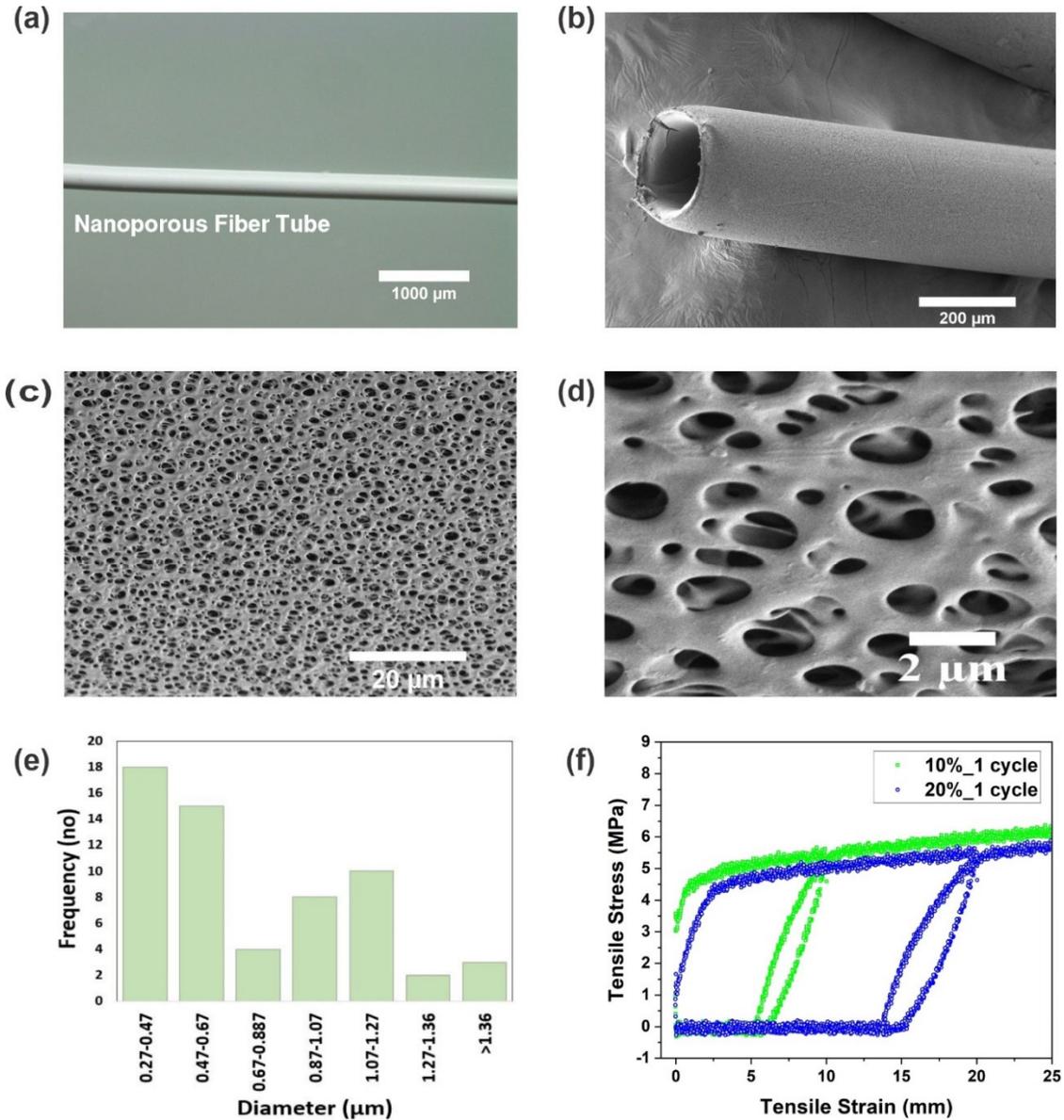

**Fig 1**: **Characterization of a mono-lumen artificial kidney** (polysulfone polymer) collected from the decades-old unused and expired dialyzer. (a) Optical image of the fiber, (b)-(d) a series of SEM images at different magnification to demonstrate the surface topography, (e) distribution of nanopores in (d) analyzed by ImageJ software, and (f) mechanical strength of an empty dialyzer tube at 10% and 20% repeated single cycling test.

After two decades of manufacturing, the mechanical strength of the strands might deteriorate and impede the upcycling process envisioned initially in this work. **Fig 1f** demonstrates the mechanical behavior of a single strand fiber under two different (10% and 20%) one-cycle loading conditions. We stretched the fiber within 1-2mm for both cycles in the linear elastic regime and 5-15mm without rupturing it entirely. Therefore, the strands that we collected have enough micro/nanopores and adequate mechanical strength (~4MPa) to hold fluid if brought in contact at the top of the cylindrical surface.

Taking inspiration from **Fig 1**, we explored conductive fluids that could be injected into these fibers to fabricate low-cost, conductive, and nanoporous wires for functional applications. We focused on Gallium alloys because of their higher surface tension, low viscosity, and surface oxide that readily forms in the air. While the research on recycling soft metals (i.e., Gallium and its alloys) has not yet gained significant interest[11]; Gallium (and alloys) as a by-product of Aluminum, Zinc, and Copper mining industries are slowly becoming critical metals for the future and imposing risks to impede the research and development progress of soft and functional materials. In the last decade, Gallium and Gallium-alloys have gained significant interest in functional studies, including soft devices, sensors, robotics, drug delivery, and shape-reconfigurable systems[6,12]. Eutectic Gallium Indium (eGaIn), as one of the most popular alloys of Gallium, has gained substantial attention in strain-responsive applications in recent years due to its fluidity, tunable conductivity, and chemically active surface oxide[13]. In the presence of air, the surface of eGaIn[14] is coated with 5nm thick Gallium Oxide ($Ga_2O_3$), allowing the bulk metal to be micro molded on-demand[6]. The oxide-enabled interfacial activities have been investigated from different viewpoints in literature where bulk metal is brought in contact with a variety of external stimuli[6]. However, less-significant efforts have been devoted to harnessing the abilities of these alloys to interact with liquids in confined domains, especially in nanoconfined spaces/pores. We believe this is because- (a) nanoporous microtubes are not readily available commercially, (b) fabrication of nanoporous microtubes is costly. To connect the two disparate problems (i.e., dialyzer recycling and Gallium reuse) and solve a global "medical waste" issue through nanotechnology, we explored several nanoporous microtubes and pneumatically injected eGaIn to upcycle artificial kidneys as the next-generation of soft conductive wires for nano/biosensing.

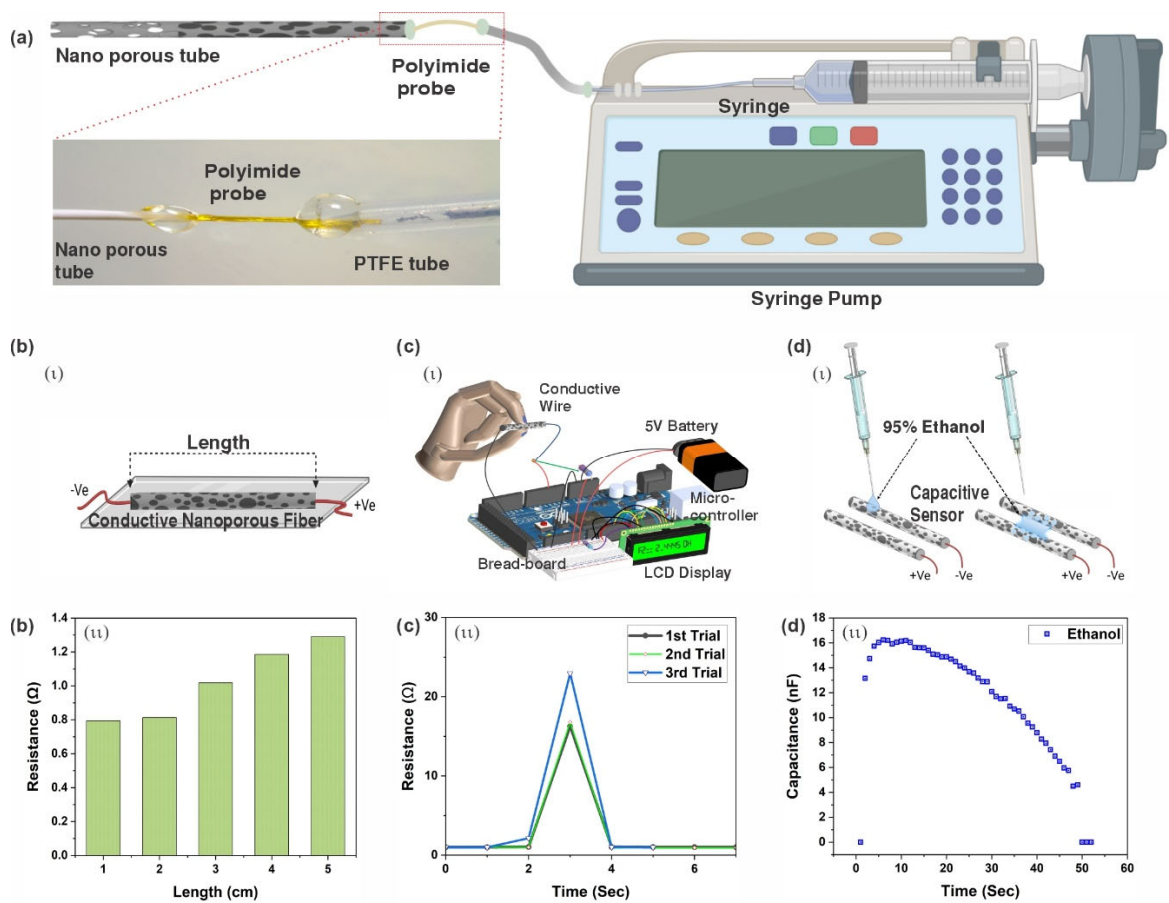

**Fig 2: Conductive artificial kidney fabrication.** (a) Pneumatically injecting eGaIn into hollow nanoporous fibers produces conductive nanoporous fibers. A custom fluidic adapter was designed (inset of a) utilizing ~120µm polyimide tubing that fits within the ~260µm polysulfone and ~850µm polytetrafluoroethylene (PTFE) tubes are bonded at two different ends using a UV glue (Bondic, Amazon). PTFE tube was connected with the metal-filled syringe on the syringe pump. [b(i-ii)] Resistance of conductive nanoporous fibers at different lengths that further utilized as a resistive sensor in an integrated circuit in [c(i)-c(ii)]. [d(i-ii)] Two conductive wires are placed side by side to form a parallel-plate capacitor to sense the capacitive response of a ~2µL EtOH droplet.

We hypothesize that injecting eGaIn into nanoporous empty fibers produces porous conductive wires. **Fig 2** demonstrates a straightforward approach of injecting eGaIn into nanoporous hollow fibers to fabricate conductive nanoporous wires (and, therefore, conductive artificial kidneys) that can emerge as new types of porous and soft sensors. Among several evident approaches of soft metallic device fabrication[15,16], such as stencil printing[17], laser patterning[18], selective wetting/dewetting[19], transfer printing[20], lithography[21], injecting eGaIn into nanoporous fibers at room temperature offers (a) the most straightforward routes to fabricate nanoporous conductive wires, (b) least-expensive metal-oxide/metal platforms to complement current-state-of-research for nano/biosensing, and (c) an exponential expansion in the research opportunities to analyze microdroplets of a variety of liquids without requiring costly tools/techniques. For the fabrication, eGaIn filled syringe mounted on a pressure-controlled syringe pump is connected with a fluidic adapter (polyimide tube in Figure 2a) through the PTFE tubing to complete the fluidic connections with the nanoporous empty fibers **[Fig 2a]**. During the pressure-driven injection process, eGaIn does not leak out of the nanopores since (a) the bulk metal flows more readily through the capillary than the nanopores, (b) the surface oxide (i.e., Gallium Oxide ($Ga_2O_3$)) that readily forms underneath the pores provides Laplace stability of the bulk metal, and (c) the pressure required to rupture the surface oxide of eGaIn through the nanopores is unattainable without damaging the pores.

The syringe pump in **Fig 2a** can deliver a maximum pressure of 1000 mbar. Our analyses on Laplace pressure[22] calculation (see supplementary document) suggest that ~39 mbar pressure requirement to induce eGaIn flow through the capillary and the minimum pressure required to flow the metal through the nanopores is ~9800 mbar. Therefore, the surface oxide provides Laplace stability of the bulk metal at the nanopores/air interface. We applied ~315-440 mbar pressure to produce conductive wires at different lengths during experiments, as shown in **Fig 2b(i-ii)**. Conductive nanoporous wires provide us an opportunity to demonstrate a touch-sensitive resistive wire [**Fig 2c (i-ii)**] and a capacitive probe to sense EtOH [**Fig 2d (i-ii)**]. While the resistive touch sensors demonstration in **Fig 2(c)(ii)** utilizing alloys of Gallium mimics touch-sensor demonstrations present in literature[23], direct acquisition of capacitive signals through interfacial interaction with microliter liquid droplets at the nanopores is the critical discovery of the present work. Prior literature findings suggest that most of the conductive wire demonstrations of eGaIn were either done on non-porous capillary or bulk metal in the presence of external stimuli[23,24]. Also, the nanoporous conductive fibers confining the $Ga_2O_3$/eGaIn interface are underexplored, and the electrical behavior of the nanoporous conductive fiber in the presence of analytical liquid droplets is still unknown.

We confirm the oxide interface underneath nanopores utilizing energy dispersive spectroscopic analyses as shown in the supporting **Fig S1.** However, the rise and drop in capacitive signal need investigation, as shown in **Fig 2d(i-ii)**. We hypothesize that, in the presence of the analytical liquid droplets, the air-$Ga_2O_3$-eGaIn interface at the nanopores changes to the liquid-$Ga_2O_3$-eGaIn interface. Also, the droplet envelopes the $Ga_2O_3$-eGaIn interface with an additional wet layer similar to the slip layer explored before[25]. We deposited a ~2µL liquid droplet (red dye mixed with DI water) on top of the nanoporous fiber and acquired images every second to further explore this mechanism. **Fig 3** depicts the dynamics of droplets on a conductive fiber. At the beginning **[Fig 3(a-f)]**, when the droplet touches the surface, fluid transport initiates through the pores —secondly, the fluid wets the entire surface along the length of the tube. The associated velocity plot has been shown in **Fig 3(g)**. We repeated these experiments for DI water, 1M HCl, and 95% EtOH and captured the velocity plots as shown in **Fig 3(h).**

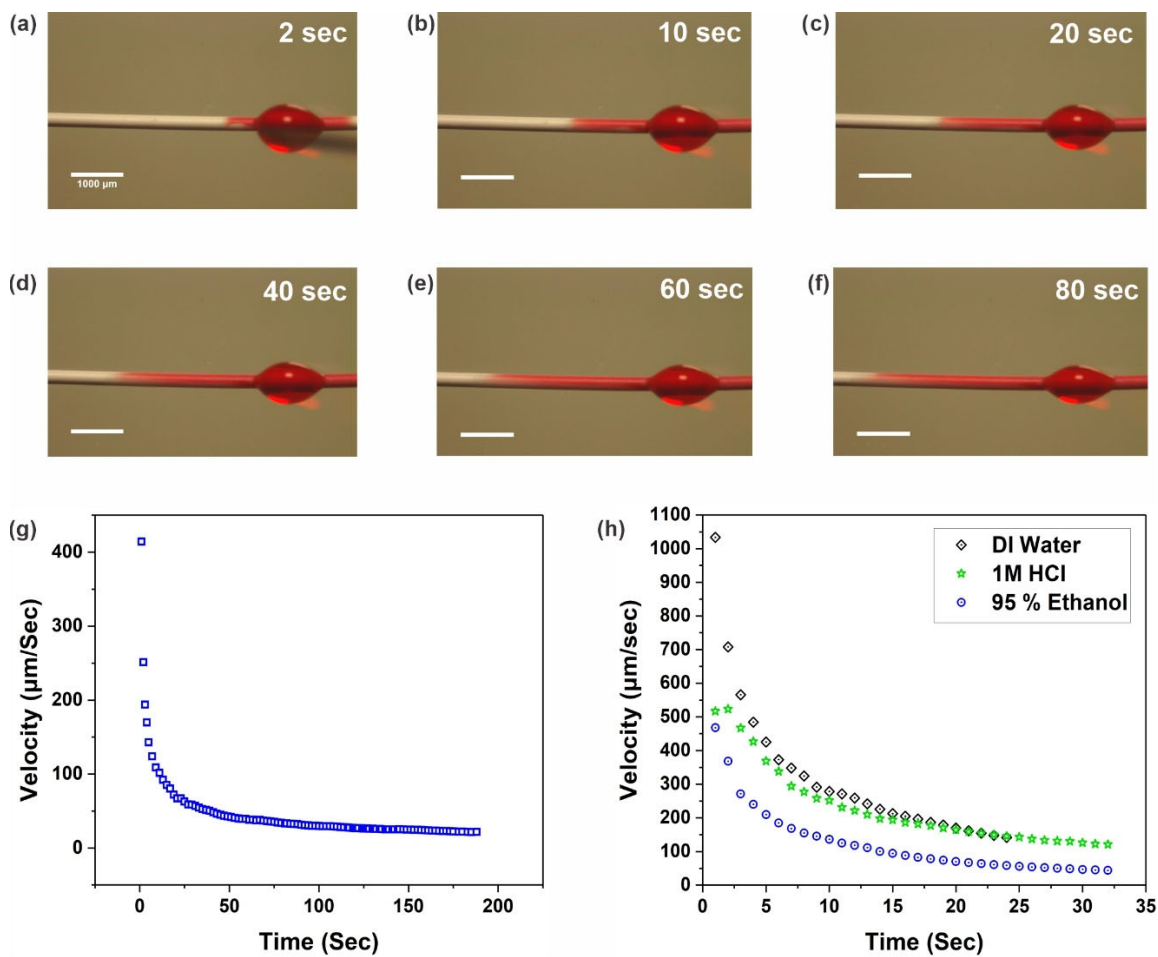

**Fig 3: Droplet dynamics on nanoporous conductive wires.** (a)-(f) a series of acquired images of red dye (food color) droplet movement on top of the nanoporous fiber to better visualize and represent the dynamic mechanism, (g) velocity decay of the dye droplet with the progression of time, (h) three different velocity patterns achieved through dispensing 2µL DI water, 1M HCl and 95% EtOH droplet on top of the sensor.

Initial velocity irrespective of the test liquids starts with a sharp spike followed by decay, suggesting capillary transport through the nanopores at the beginning. As the height of the droplet is shortened (i.e., change in the radius of curvature of the droplets), capillary tension diminishes, but the fluid continues to spread and wet the surface of the conductive nanoporous wires. If the capillary transport is followed by a diffusive transport of the analytes, then the molecular weight of the analyte should govern the flow physics of microdroplets[26]. **Fig 3h** indicates that as we change liquids from DI water to EtOH, the molecular size gets more prominent, and the velocity due to diffusive transport goes down. Also, our maximum velocity corresponds to the scaling laws for diffusive transport (i.e., velocity ~ (length/time) ~ $\frac{\sqrt{Diffusivity}}{\sqrt{time}}$)[26]. It further validates the working principle of fluid transport on nanoporous conductive wires. It is worth mentioning that we did not treat the surface of the conductive nanopores with any chemicals/vapors. As soon as we bring the droplets on the fibers, the initial states are shown in **Fig S2.** The lifetime of a water droplet (i.e., the time it takes to spread) is higher than that of 1M HCl and 95% EtOH due to the initial hydrophobicity of the polysulfone as shown in **Fig. S3;** but the droplet eventually spreads due to capillarity. Therefore, fluids with lower surface energy can spread faster and wet the conductive nanoporous fiber like EtOH.

The spreading time and the wetting dynamics further motivated us to explore two demonstrations where conductive nanoporous fibers could be harnessed as low-cost, low-power and lightweight sensors in an integrated circuit platform. **Figs. 4a** and **4c** are examples of the conductive nanoporous fibers as capacitive sensors to harness microdroplets and liquid levels. The capacitive sensors can detect the presence of the liquid as soon as fluids touch the capacitive wires, and the circuit ceases as the liquid film breaks from the wires. We assume that in a parallel plate capacitor, if the distance between two identical (i.e., equal length and initial state) conductive nanoporous wires is kept constant, the initial capacitance in the air can be perturbed by the presence of the liquids. Also, the concentration-dependent dielectric permittivity of each liquid influences the capacitance[27].

We dispensed ~2μL droplets of DI water, 1M HCl, and 95% EtOH between two conductive wires acted as parallel plate capacitors in an integrated circuit, as shown in **Fig 4a**. The corresponding change in capacitances from the baseline values are given in **Fig 4b**. For the droplet of DI water, we noticed the residence time of the liquid film touching both plates was the longest, and for 95% EtOH, and 1M HCl, the times were almost similar. Also, for 1M HCl and EtOH, we noticed a rapid increase in capacitance compared to the DI water droplet. Though the dielectric permittivity of water[28] is most prominent among the three droplets, it alone cannot explain the capacitive response that we noticed in **Fig 4b** because of the wetting dynamics and possible shift in a new oxide at the interface of eGaIn than native oxide[25]. The etching chemistry of eGaIn with HCl and interaction behavior in water and alcohol may have further influenced the results we obtained[6,29]. The presence of the bare metal in the case of HCl; and a new surface oxide in the case of water/alcohol may also have contributed to the rise in capacitance. Dipping the parallel plates into a beaker partially filled with DI water allowed us to demonstrate a liquid level sensor, as shown in **Fig 4c**. We noticed that the change in capacitance varied by the % of initial length dipped into the DI water, as shown in **Fig 4d**. Though these observations demonstrate a new class of soft materials, we believe the fluid-eGaIn interaction at the nanopores opens wider research avenues that literature lack currently.

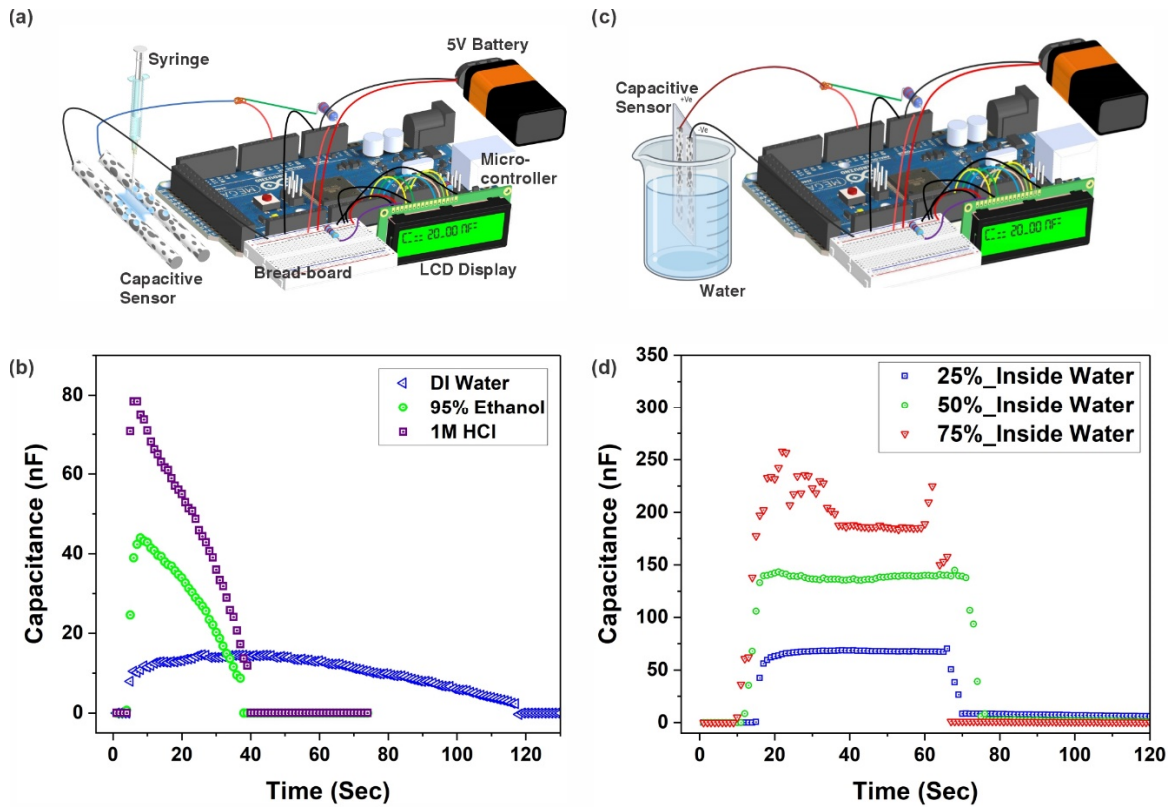

**Fig 4: Proof-of-concept capacitive sensor demonstrations.** (a) Conductive nanoporous fibers as a capacitive sensor in an integrated circuit which facilitates sensing of test droplet, (b) Three different capacitive response achieved while 2μL DI water, 1M HCl and 95% EtOH droplet dispensed on top of the sensor, (c) An integrated circuit of parallel plate capacitive sensor placed on a glass slide for sensing water level, (d) Change in capacitance for dipping the parallel plate capacitive sensor into DI water at three different heights.

Here, we revealed a new class of conductive and nanoporous functional materials utilizing eGaIn in expired artificial kidneys from dialyzers. As a proof-of-concept and first-of-its-kinds system, we also presented both the resistive and capacitive behavior of the conductive wires in the presence of external touch and microliter droplets of analytical liquids. We also showed the dynamics of droplets on the surface of the conductive wires and how the dynamics could be further explored as the capacitive wires to sense and detect known analytes (i.e., DI water, 1M HCl, and EtOH) and liquid levels. The current literature lacks knowledge on eGaIn-based nanoporous devices that can respond to a biomolecule (e.g., water and EtOH) and where the nanopores can influence the dynamic motion of different fluids differently. Also, the surface oxide of eGaIn is in contact with millions of micro/nanopores [cf. see supporting document for pore count] of the conductive wires increases the active $Ga_2O_3$/eGaIn interfaces at each nanoconfinement. A series of future explorations could potentially unlock the utilities of such systems in other applications, i.e., sensing, fluidics, dilute sample analyses, recycling of soft metal alloys, medical waste upcycling, and energy harvesting.

**Acknowledgements:**

We thank Prof. Chuck Coronella, Department of Chemical and Materials Engineering, University of Nevada, Reno for providing the dialyzer. Dr. Khan acknowledges the financial support received from the VPRI's startup package, University of Nevada, Reno; Syringe grant support from Hamilton Robotics Inc; Differential fees award support from College of Engineering, University of Nevada, Reno.

**Contributions:**

R.K designed and conceived the project. MM designed, experimented, and analyzed data. GL experimented. RK and MM edited, and revised drafts, GL reviewed.

**Corresponding Author**

**Dr. M. Rashed Khan, mrkhan@unr.edu**

## Method
### Imaging and Characterization

Initially, we took ~3.5 cm long polysulfone nanofiber (NF) tube from twenty years old Hemoflow F80A dialyzer. The image of the NF tube was taken under Stereomicroscope (OLYMPUS SZ61), and the image was recorded using AmScope software (**Fig 1a**). To capture the surface topography of the NF tube, a scanning electron microscope (SEM) was used. Several ~2mm length of NF tube was cut and placed on the SEM sample holder using double-sided carbon tape. Then it was sputter-coated with gold for 10 sec to increase SEM image contrast and avoid charging problems. Then images of the empty NF tube were taken under SEM (**Fig 1b**). In this case, the sample was placed on a 45° uphill holder. The imaging parameters were: magnification: 200x, high voltage: 5 kV, current: 0.4 nA, mode: scanning electron (SE), working distance: 3.8 mm. In **Fig 1c,** the sample was placed on a flat holder. The imaging parameters were: magnification: 5000x, high voltage: 5 kV, current: 0.4 nA, mode: scanning electron (SE), working distance: 3.0 mm. **Fig 1d** was obtained by taking the image under SEM at 20000x magnification, 5 kV high voltage, 0.4nA current, scanning electron (SE) mode, and 3.0 mm working distance. **Fig 1d** was analyzed using the ImageJ software. During this analysis, the data for the pore diameter distribution was collected. Then it was plotted in excel, shown in **Fig 1e**. To determine the mechanical strength of the tube, tensile stress-strain data for the NF tube was collected. This experiment was done using Instron 5982 for a single cycle test. The geometry was selected as Tubular, Initial Length was 100mm, and the strain rate was 50 mm/min.

### Fabrication Method

The liquid metal-filled tube was fabricated utilizing a pressure-controlled microfluidic pump Fluigent by quantifying the Laplace pressure for a cylindrical surface. The critical pressure is inversely related to the channel radius (see supporting document). So, the narrow channels require higher pressure than channels with wider cross-sections. In the case of nanoporous fiber tubes, the tube radius is 130 μm, and the surface tension of EGaIn is 500 mN/meter. To inject the liquid metal into the tube, a minimum pressure of 38.46 mbar is required. However, we prepared a setup to ensure smooth injection of liquid metal into the nanofiber tube to test this. A small thermoplastic elastomer tube allows the liquid metal transfer from the reservoir through the polytetrafluoroethylene (PTFE) tube utilizing an automated pressure-driven pumping system. This approach can be combined with our pressure control micropump, using a syringe as the master reservoir filled with EGaIn. One end of a small PTFE tube was connected with the needle tip of the reservoir, and the other end was connected with the microlumen Polyimide (MLP) probe of around 218 μm diameters. The other end of the MLP probe was injected into the nanofiber tube close to the entrance of a loading channel. The use of a low diameter capillary considerably reduces sample consumption and facilitates the fluid flow without requiring flow at unwanted areas, which can, in some cases, lead to bubble nucleation or bending of the delicate fiber tube. The tube and probe connections were sealed using UV glue (Bondic, Amazon Inc.). This technique facilitates the flow of quantities on the sub-microliter scale into channels having an inner diameter in the micron range while keeping a large volume reservoir filled with liquid metal. We applied 315-440 mbar pressure to fill the tube; however, we did not see any metal leaks through the nanopores. We calculated the theoretical pressure required and found that it was ~10000 mbar to come out of the nanopores. The metal, therefore, experiences less pressure flowing through the tube than the nanopores. While finishing the injection of liquid metal, backflow through the injection channel is supposed to occur. So, the pressure should be released slowly. **Fig 2a** shows the depiction of the liquid metal-filled tube fabrication system.

**Testing Touch Response**

After successfully injecting the liquid metal inside the tube, we inserted copper wires at the two ends of the sensor, and the connection was sealed using UV glue. To test the conductivity of this electrode, five different lengths of 5cm, 4cm, 3cm, 2cm, 1cm LM filled NT were prepared. The initial resistivity of these sensors was measured using a multimeter (Keysight-U1733C). This has been depicted in **Fig 2b**. Later, the resistive change of the sensor was observed in response to applied pressure. Only by squeezing the volume of the liquid metal a change in the resistive signal be observed. The sensor was pressed three different times, and the change in resistance was recorded using a multimeter (Keysight-U1733C). **Fig 2c** depicts the touch-sensitive response of the system.

**Capacitive Response**

To analyze another functional property, we bring two parallel plates nearby; one is connected to the positive terminal, and the other end is grounded to mimic a parallel plate capacitor. When we bring alcohol molecules, we find a significant response of capacitance. We continuously tracked the response using a multimeter (Keysight-U1733C), and it has been shown in **Fig 2d**.

**Microfluidic Behavior**

A sensor was placed under the Stereomicroscope (OLYMPUS SZ61). A Dye-water droplet was placed on top of the sensor, and a continuous image at a 1sec interval was recorded using AmScope software. Later these images were analyzed using ImageJ software. The velocity of the droplet was extracted and plotted using Origin software, as shown in **Fig 3g**. A similar experiment was done for a 2 μL drop of DI water, 1M HCl, and 95% Ethanol. These drops were harnessed on the sensor, respectively. Furthermore, the images of the fluid flow were taken from the top using a stereoscope. The velocity data were plotted using Origin software shown in **Fig 3h**.

**Sensor Applications**

In the first demonstration of the capacitive response, the distance between the two parallel eGaIn filled tube was fixed at 1mm. Then 2μL of DI water was applied to the sensor, and the response was recorded. This was repeated for 95% Ethanol and 1M HCl. The three responses were continuously tracked using a multimeter (Keysight-U1733C) and plotted (**Fig 4b**) using Origin software. It is known that every liquid solution has a specific dielectric permittivity. Dielectric permittivity ($\varepsilon$) is the ability of a substance to hold an electrical charge. It should also vary from molecule to molecule. Besides, the wetting dynamics and possible shift in a new oxide at the interface of eGaIn also impacted the response. For the fluid level sensing application, shown in **Fig 4c**, the sensor was placed on a glass slide, and then the lens length was divided into four sections and marked on the glass slide. Then 25% of the sensor was dipped inside water and held there for 1min, and the response was recorded. Similarly, 50% and 75% of the sensor were dipped, and the response was recorded. Then these responses were continuously tracked using a multimeter (Keysight-U1733C) and plotted (**Fig 4d**) using Origin software.

# Supplementary Information
## Confining Eutectic Gallium Indium (eGaIn) in Expired Artificial Kidneys to Unveil Nanoporous Conductive Wires

Momena Monwar[a], Gerra Licup[a], M. Rashed Khan[a,*]

[a]Department of Chemical and Materials Engineering, University of Nevada, Reno, NV-89557

[*]Corresponding Author: mrkhan@unr.edu

**Quantification of the Applied Pressure during Fabrication**

Our fabrication method involves the injection of the liquid metal using a pressure-controlled microfluidic pump by quantifying the Laplace pressure that harnesses the unique rheological properties of eGaIn[1].

Using this concept, we quantify the applied pressure to induce the metal flow through the nanoporous hollow capillary[2] using equation (i). The Laplace pressure $\triangle P$ of the liquid in the channel can be calculated according to the Laplace equation:

$$\triangle P = \gamma \left(\frac{1}{R_1} + \frac{1}{R_2}\right) \quad \text{.......................... (i)}$$

Here $\gamma$ is the interfacial surface tension of eGaIn), $R_1$ and $R_2$ are the principal radii of the curvature.

We applied 315-440 mbar pressure to fill the tube; however, we did not see any metal leaks through the nanopores due to the unattainable high pressure at the nanopore.

**Oxide Interface Underneath Nanopores:**

We performed energy dispersive spectroscopy (EDS) analysis to determine elemental compositions in the nanopores. We performed the EDS of an empty nanofiber tube and a liquid metal-filled nanofiber tube. The oxide percentage for the liquid metal-filled tube was higher than the empty one. Also, there was the presence of Gallium and Indium at a significant amount. **Fig S1(a)-(b)** shows the EDS spectrum obtained for empty (spectrum 83) and liquid metal-filled nanofiber tube (spectrum 103), respectively. We also investigated a single nanopore of the sensor and observed the presence of oxide at a significant amount due to the chemical composition of polysulfone. **Fig S1(c)** represents the spectrum 109 for a single nanopore.

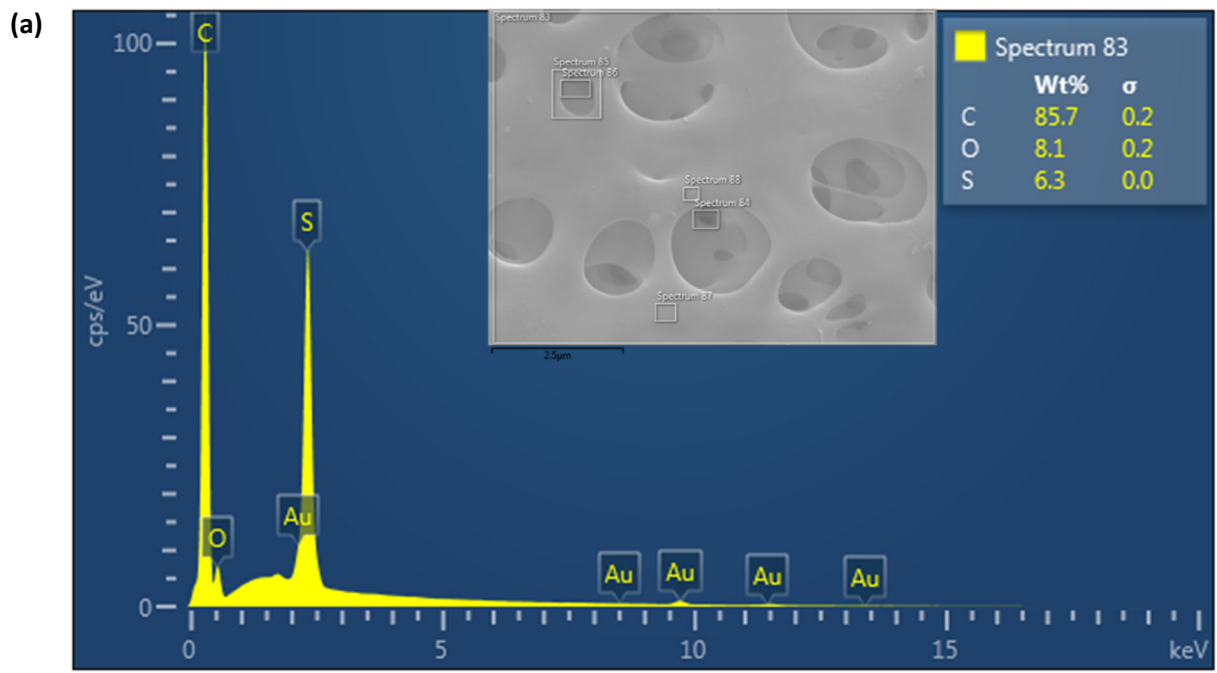

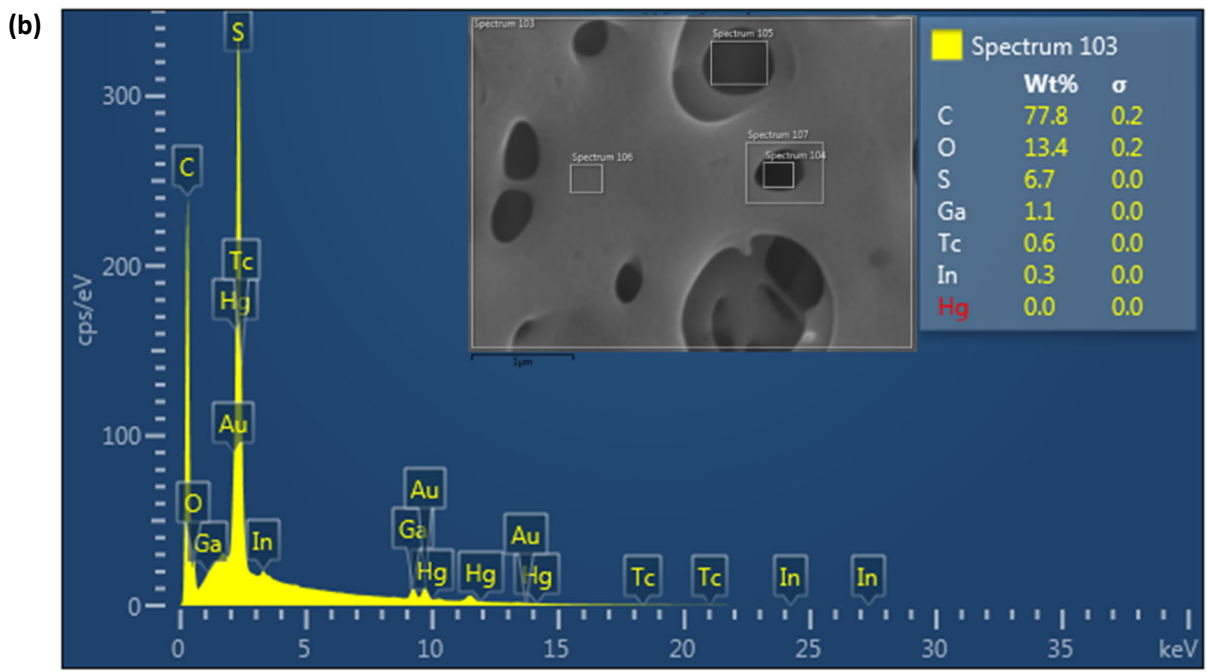

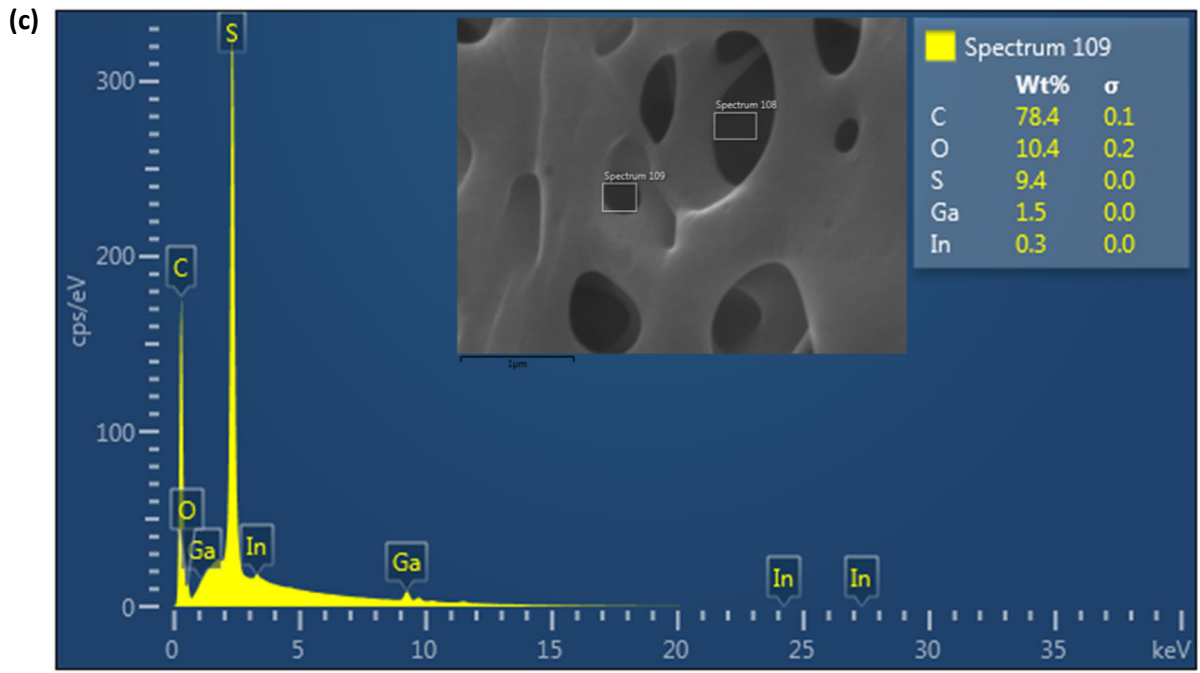

**Fig S1: EDS Spectrum.** (a) presence of elements onto and underneath the empty nanofiber tube surface, (b) spectrum indicating the percentage of elements present onto and underneath the sensor, (c) spectrum obtained from a single nanopore reflecting the oxide interface underneath nanopores.

**Pore calculation:**

This section assumes the total possible number of pores underneath a 2µL droplet of test fluid. We took some known areas with a known number of pores to predict the pore covered by the droplet.

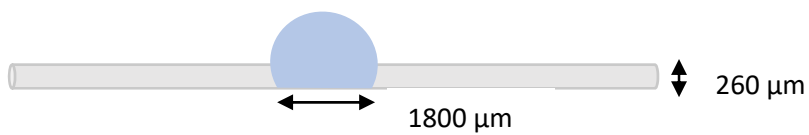

Surface area covered by the droplet, $A_1 = 2\pi r(r + h) = 1.58$ mm$^2$

Where, r=130 µm, h= 1800 µm

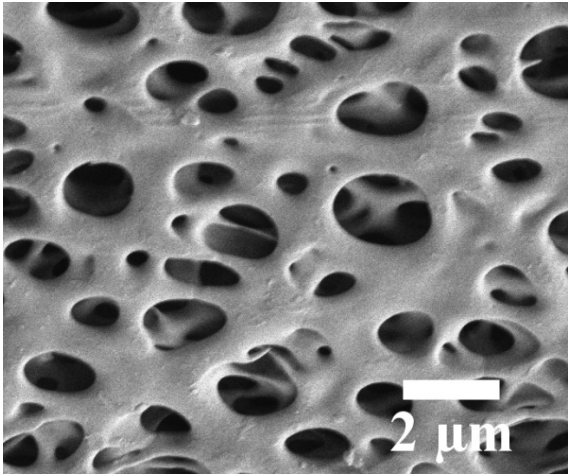

Image area 1:

53 count pore

Area= (12.209×9.93)=121.24 µm² =0.00012124 mm²

Predicted pore covered by the droplet= $\frac{1.58 \times 53}{0.00012124}$ = **690,696 Count**

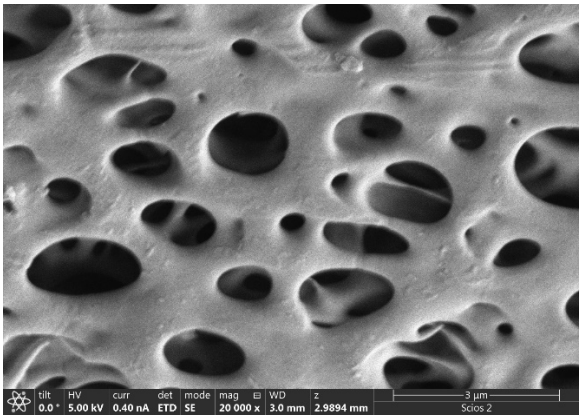

Image area 2:

33 count pore

Area= (10.425×6.938)=72.33 µm² =0.00007233 mm²

Predicted pore covered by the droplet= $\frac{1.58 \times 33}{0.00007233}$ = **720,862 Count**

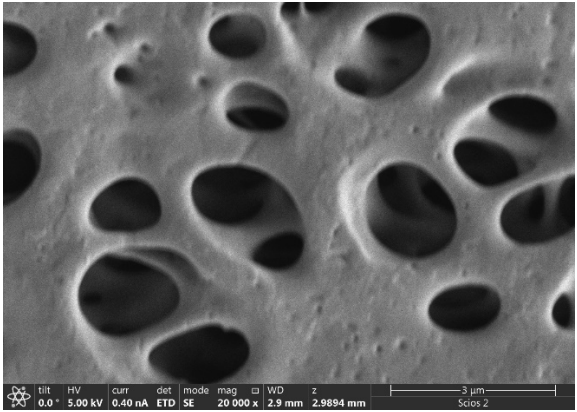

Image area 3

Image area 3:

17 count pore

Area= (10.385×6.937)=72.04 µm² =0.00007204 mm²

Predicted pore covered by the droplet= $\frac{1.58 \times 17}{0.00007204}$ = **372,844 Count**

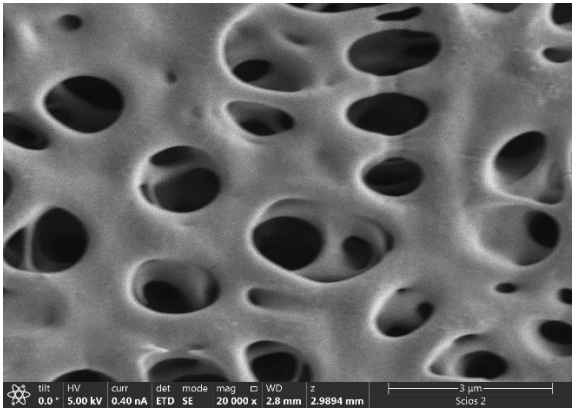

Image area 4

Image area 4:

23 count pore

Area= (10.426×6.964)=72.61 µm² =0.00007261 mm²

Predicted pore covered by the droplet= $\frac{1.58 \times 23}{0.00007261}$ = **500,505 Count**

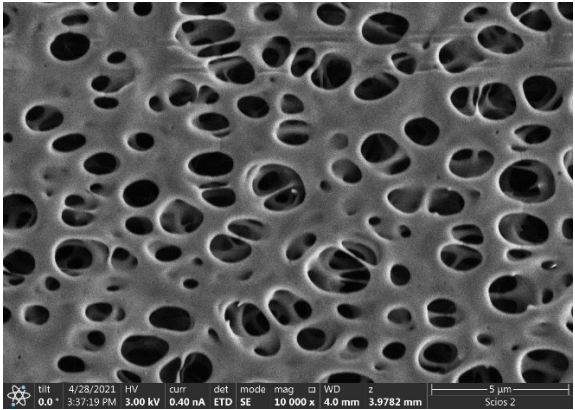

Image area 5:

98 count pore

Area= (20.984×13.962)=292.98 µm² =0.00029298 mm²

Predicted pore covered by the droplet= $\frac{1.58 \times 98}{0.00029298}$ = **528,500 Count**

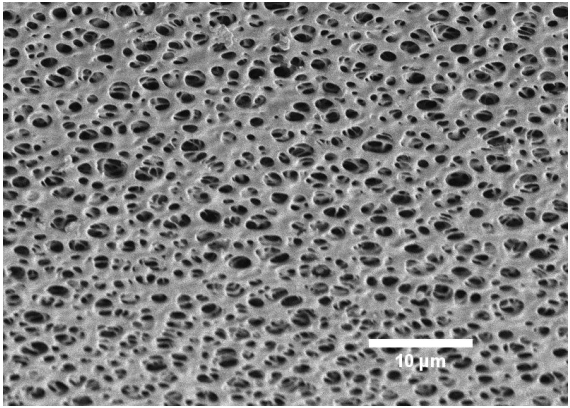

Image area 6:

666 count pore

Area= (54.231×38.825)=2105.52 µm² =0.00210552 mm²

Predicted pore covered by the droplet= $\frac{1.58 \times 666}{0.00210552}$ = **499,772 Count**

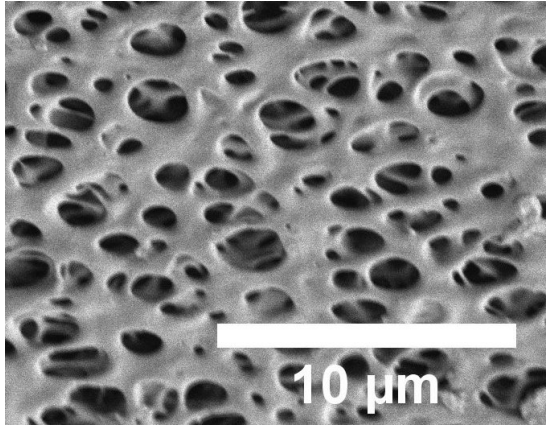

Image area 7

Image area 7:

93 count pore

Area= (18.064×14.167)=255.912 µm² =0.000255912 mm²

Predicted pore covered by the droplet= $\frac{1.58 \times 93}{0.000255912}$ = **574,181 Count**

**State of Droplets at the different time frame and Droplet lifetime:**

For observation of the droplet lifetime, a nanoporous conductive sensor was placed in front of the goniometer (rame-hart, 500-U-2) camera, and then a 2µL droplet of DI water was placed on to the sensor, and continuous images at 1-sec intervals were taken. Then the lifetime of the droplet was analyzed from the images. This same experiment was repeated for 1M HCl and 95% Ethanol.

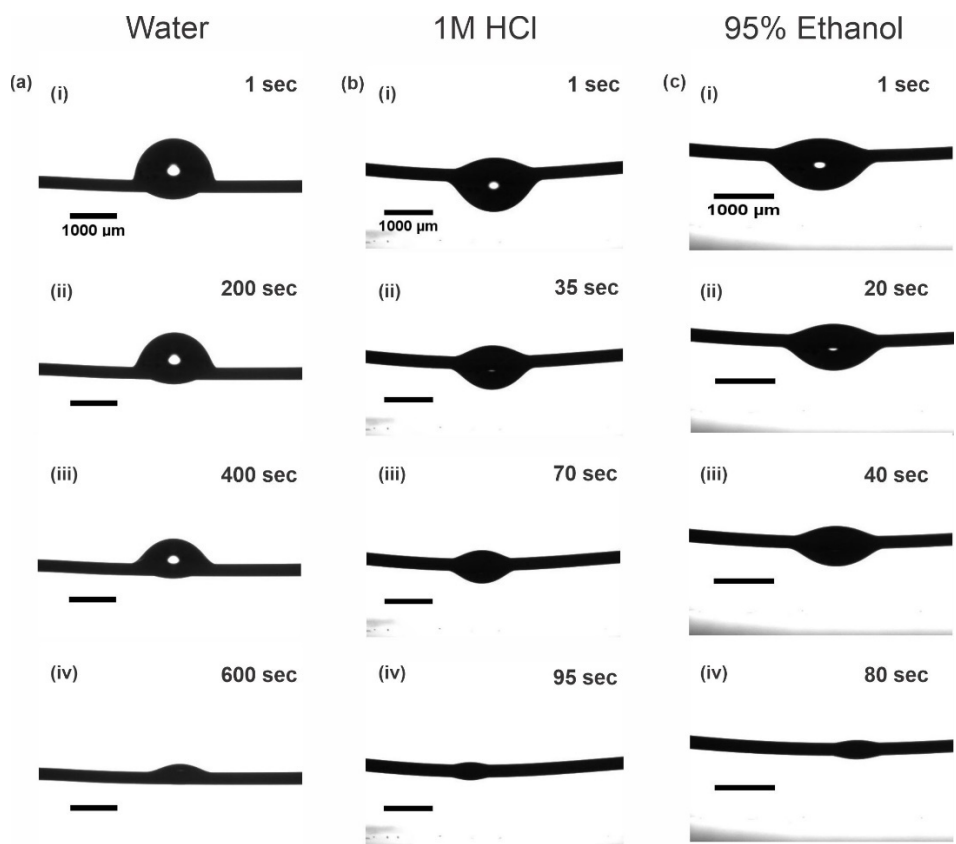

**Fig S2: Droplet lifetime.** [a(i-iv)] images of progression of 2µL DI water droplet on sensor surface, [b(i-iv)] images of progression of 2µL 1M HCl droplet on sensor surface, [c(i-iv)] images of progression of 2µL 95% EtOH droplet on sensor surface.

**The contact angle of water droplet on sensor surface:**

To perform the contact angle measurement, we placed a conductive nanoporous sensor in front of the goniometer (rame-hart, 500-U-2) camera, and then a 2µL droplet of DI water was placed on to the sensor, and continuous images at 1-sec intervals were taken. The Contact angle was obtained from the drop image software. The contact angle data was plotted using origin software.

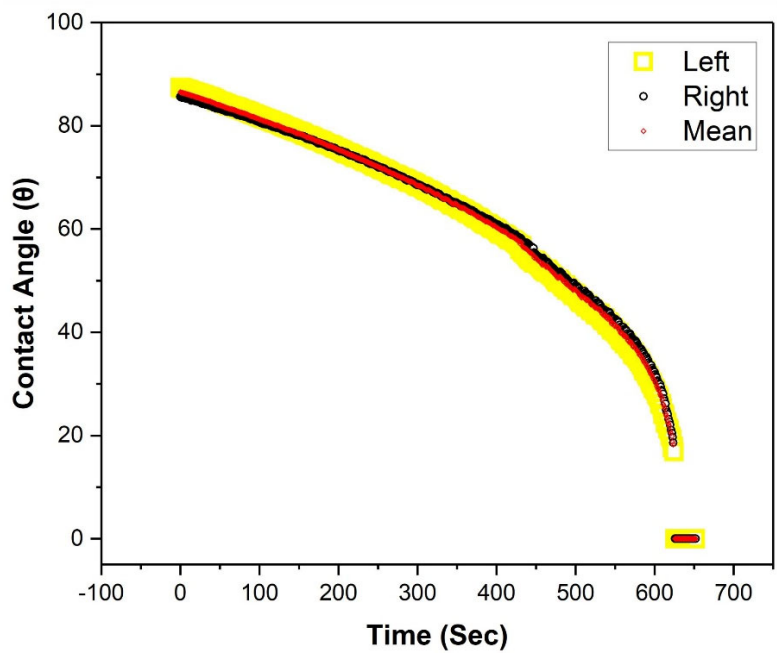

**Fig. S3: Contact angle.** Change in contact angle of ~2 µL water droplet on the nanoporous conductive wire. The initial contact angle demonstrates hydrophobicity; however, the diminishing angles due to capillary transport of fluids lower the contact angles.

**Verification of the volume of test droplets:**

We dispensed the 2µL droplets by pipetting. We also verified that the droplet amount was similar for dispensing each time. Initially, a glass slide was plasma treated first, and then it was ethoxy silane treated. After that, multiple water droplet using a 2µL pipet was dropped on a glass slide. The images and the volume experiment were performed using the goniometer (rame-hart, 500-U-2). The data of the droplet volume was obtained from the drop image software. The droplet volume data was plotted using origin software.

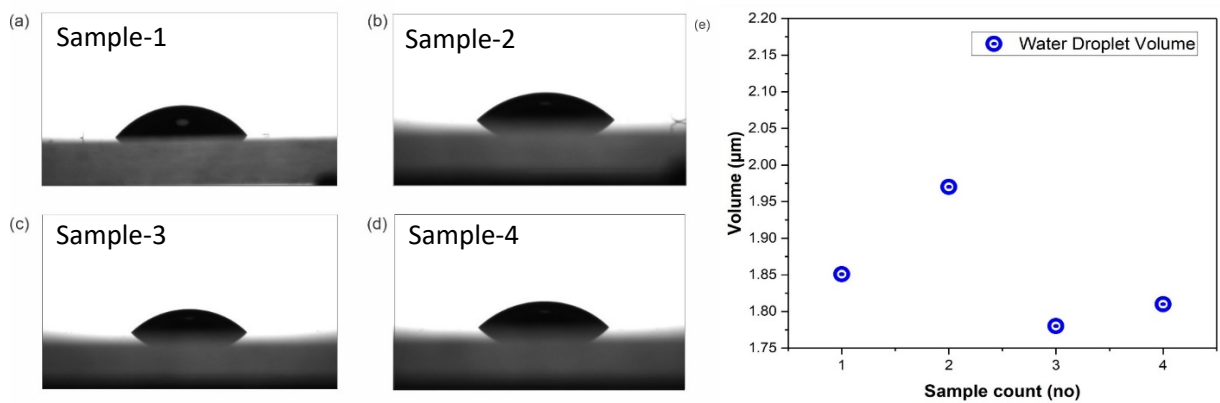

**Fig S4: Droplet volume.** (a)-(d) images of a series of water droplets dispensed using a 2μL pipette using a goniometer, (e) acquired volume of the sample water droplets.

**Pore diameter distribution details:**

Table I: This diameter data were collected from **Fig 1d** of the main manuscript using ImageJ software

| Diameter (nm) | | | |
|---|---|---|---|
| 273 | 445 | 623 | 1089 |
| 304 | 462 | 628 | 1101 |
| 316 | 470 | 633 | 1103 |
| 327 | 474 | 709 | 1109 |
| 338 | 481 | 734 | 1127 |
| 344 | 485 | 836 | 1143 |
| 362 | 487 | 861 | 1170 |
| 365 | 497 | 894 | 1186 |
| 368 | 506 | 978 | 1215 |
| 372 | 520 | 1000 | 1247 |
| 397 | 553 | 1039 | 1329 |
| 402 | 577 | 1043 | 1354 |
| 406 | 597 | 1047 | 1787 |
| 412 | 599 | 1051 | 1800 |
| 419 | 614 | 1053 | 2039 |

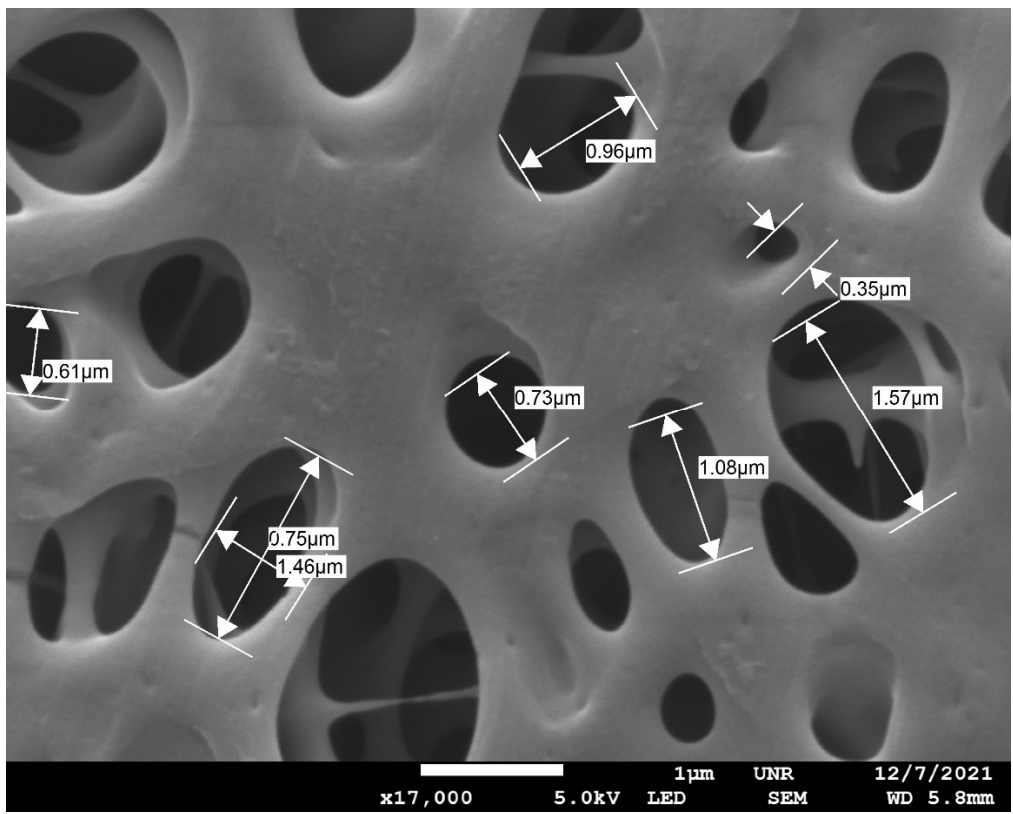

**Fig S5: Pore diameter.** Image of the pore diameter directly obtained from the SEM software.

**Droplet Lifetime Videos:**

The associated droplet lifetime videos are uploaded along with this supporting document. 'Water_Droplet_Lifetime' is the video of 2μL DI water droplet progression on the sensor surface. 'HCl_Droplet_Lifetime' is the video of 2μL 1M HCl droplet progression on the sensor surface, and 'Ethano_Droplet_Lifetime' is the video of 2μL 95% EtOH droplet progression on the sensor surface. All the videos were taken from the side with a goniometer.

**Droplet Velocity Videos:**

The associated droplet lifetime videos are uploaded along with this supporting document. 'DI Water_Droplet_Velocity' is the video of 2μL DI water droplet movement on the sensor surface. 'HCl_Droplet_Velocity' is the video of 2μL 1M HCl droplet movement on the sensor surface. 'EtOH_Droplet_Velocity' is the video of 2μL 95% EtOH droplet movement on the sensor surface. All the videos were taken from the top with a stereomicroscope.